\documentclass[aps,prc,letterpaper,11pt,twoside,tightenlines,nofootinbib,showpacs,preprint]{revtex4-1}
\usepackage{graphicx}
\usepackage[sort&compress]{natbib}
\usepackage{subfigure}
\usepackage{amsmath}
\usepackage{amsfonts}
\usepackage{cancel}
\usepackage{amssymb}
\usepackage{hyperref}
\begin{document}
\arraycolsep1.5pt
\newcommand{\Ima}{\textrm{Im}}
\newcommand{\Rea}{\textrm{Re}}
\newcommand{\mev}{\textrm{ MeV}}
\newcommand{\be}{\begin{equation}}
\newcommand{\ee}{\end{equation}}
\newcommand{\ba}{\begin{eqnarray}}
\newcommand{\ea}{\end{eqnarray}}
\newcommand{\gev}{\textrm{ GeV}}
\newcommand{\nn}{{\nonumber}}
\newcommand{\dtres}{d^{\hspace{0.1mm} 3}\hspace{-0.5mm}}
\newcommand{\rts}{ \sqrt s}
\newcommand{\non}{\nonumber \\[2mm]}

\title{\boldmath$f_1(1285)$ decays into $a_0(980)\pi^0$, $f_0(980)\pi^0$ and isospin breaking}

\author{F. Aceti$^{1}$, J. M. Dias$^{1,2}$ and E. Oset$^{1}$}
\affiliation{$^{1}$Departamento de F\'{\i}sica Te\'orica, Universidad de Valencia and IFIC, Centro Mixto Universidad de
Valencia-CSIC, Institutos de Investigaci\'on de Paterna, Aptdo. 22085, 46071 Valencia,
Spain\\ \\
$^{2}$Instituto de F\'isica, Universidade de S\~ao Paulo, C. P. 66318, 05389-970, S\~ao Paulo, SP, Brasil.
 }

\date{\today}

\begin{abstract}
We evaluate the decay width for the processes $f_1(1285) \to \pi^0 a_0(980)$ and $f_1(1285) \to \pi^0 f_0(980)$ taking into account that all three resonances are dynamically generated from the meson-meson interaction, the $f_1(1285)$ from $K^* \bar K -c.c$ and the $a_0(980)$, $f_0(980)$ from $\pi \eta, K \bar K$ and $\pi \pi, K \bar K$ respectively.  We use a triangular mechanism similar to that of the $\eta(1405) \to \pi \pi \eta$, which provides a decay width for $f_1(1285) \to \pi^0 a_0(980)$ with a branching fraction of the order of 30\%, in agreement with experiment. At the same time we evaluate the decay width for the isospin forbidden  $f_1(1285) \to \pi^0 f_0(980)$, which appears when we consider different masses for the charged and neutral kaons, and show that it is much more suppressed than in the $\eta(1405) \to \pi \pi \eta$ case, but gives rise to a narrow shape of the $\pi^+ \pi^-$ distribution similar to the one found in the $\eta(1405) \to \pi \pi \eta$ decay.
\end{abstract}

\maketitle
\raggedbottom

\section{Introduction}
\label{Intro}
The quest for understanding the nature of recently discovered mesons and baryons is one of the most challenging topics in hadron physics. The traditional idea that they could be described as composite $q\bar{q}$ and $qqq$ objects has been reviewed to leave room for more complex structures involving more quarks in some cases \cite{Klempt:2007cp,Klempt:2009pi}. The application of chiral dynamics to the interaction of hadrons has had a significant success in recent years. An effective approach to QCD at low energy is provided by  Chiral Lagrangians, in which mesons and baryons appear as fundamental degrees of freedom \cite{Weinberg:1968de, Gasser:1983yg, Ecker:1994gg,Bernard:1995dp}.

 However, due to its limited range of convergence, the perturbation theory derived from these Lagrangians, Chiral Perturbation Theory ($\chi PT$), proved insufficient to describe the hadron spectrum, and its non-perturbative unitary extension, called chiral unitary approach \cite{Kaiser:1995eg,Kaiser:1996js,npa, Kaiser:1998fi,ramonet,angelskaon,ollerulf,Jido:2002yz,carmen,cola,carmenjuan,hyodo,review}, became necessary. This new method allows us to explain many mesons and baryons as composite states of hadrons, among them the $a_0(980)$ and $f_0(980)$, and the $f_1(1285)$, which are considered as dynamically generated resonances. 
 
Throughout this work, we will consider these resonances as dynamically generated from the interaction of two mesons, appearing as poles in the complex plane of the scattering amplitudes. These amplitudes are obtained using the tree level meson-meson interaction potentials, derived from the chiral Lagrangians, as a kernel for the Bethe-Salpeter equation in coupled channels. The $f_1(1285)$ resonance appears, together with the block of axial vector mesons, from the pseudoscalar vector interaction in $I=0$, $J=0$, in coupled channels. In \cite{Lutz:2003fm} a speed plot of the amplitude was used to identify these resonances. In \cite{f1luis} poles of the amplitudes in the complex plane were searched for in order to identify the resonances. An extension of the work of \cite{f1luis} including higher orders in the potential is done in cite \cite{Zhou:2014ila}, where the results are also extrapolated to higher $m_{\pi}$ masses  to match results of eventual QCD lattice results. In this latter work the effect of higher orders in the potential for the $f_1(1285)$ was found negligible. Extrapolation of these ideas to the charm sector is done in \cite{Guo:2006rp}. The $f_{1}(1285)$ resonance is peculiar in the sense that it comes from a single channel $K\bar{K}^*$. On the other hand the building blocks of the $a_0$ and $f_0$ are  the meson pairs $\pi\eta$, $K\bar{K}$ and $\pi\pi$, $K\bar{K}$ respectively \cite{npa,kaiser,markushin,juanito,rios,nebreda}.

Since isospin symmetry is broken in meson rescattering due to the different masses of charged and neutral kaons, the $a_0(980)$ and $f_0(980)$ can mix, and quantifying this mixing can help understanding the nature of these two resonances. The topic was first discussed in Ref. \cite{Achasov:1979xc} and more recently in \cite{Wu:2007jh, Hanhart:2007bd}, in which the reaction $J/\psi\rightarrow \phi\pi^0\eta$ was studied. The same reaction was also studied in \cite{Roca:2012cv} using the chiral unitary approach, as in \cite{Hanhart:2007bd}. In Ref. \cite{Wu:2008hx} the reaction $\chi_{c1}\rightarrow\pi^0\pi\eta$ was also proposed as a test to measure the amount of $a_0(980)-f_0(980)$ mixing.

The reaction $J/\psi\rightarrow \phi\pi^0\eta$ was then studied at BES \cite{Ablikim:2010aa}, where a narrow signal of the order of the difference of kaon masses, in agreement with the predictions of \cite{Wu:2007jh, Hanhart:2007bd}, was found, with an intensity of half percent with respect to the one of $J/\psi\rightarrow \phi\pi^+\pi^-$ in the region of the $f_0(980)$ peak of the $\pi\pi$ distribution. The same experimental work reports also on the reaction $\chi_{c1}\rightarrow\pi^0\pi\pi$ in the region of the $f_0(980)$, finding again a very narrow signal with an intensity of half percent with respect to $\chi_{c1}\rightarrow\pi^0\pi\eta$ in the $a_0(980)$ region. These findings are within expected values for isospin violation and the narrowness of the signal is tied to the mass difference between the kaons, which results in a difference between the loop functions for the intermediate kaons in the rescattering leading to the production of the $a_0(980)$ and $f_0(980)$. These results provide support to the composite meson-meson nature of these resonances.

In a recent paper the BES team has reported a larger isospin violation in the decay $\eta(1405) \to \pi^{0} f_0(980)$ compared to the $\eta(1405) \to \pi^{0} a_0(980)$, about $18\%$ \cite{BESIII:2012aa}, very difficult to explain unless one considers the $\eta(1405)$ as a mixture of $I=0$ and $I=1$. However, this would lead to the production of the $f_0(980)$ with its natural width, around $50$ MeV, and not with the one of about $9$ MeV observed by BES. 

In Ref. \cite{wuzou}, a particular mechanism was proposed to study these processes, consisting in the $\eta(1405)$ decaying into $K^*\bar{K}$, the posterior $K^*$ decay into $\pi^0K$ and the rescattering of the $K\bar{K}$ to produce either the $f_0(980)$ and the $a_0(980)$ resonances. This leads to a triangular loop diagram that has two cuts (singularities in the integrand), which make it different from the standard G loop function from $K\bar{K}$, with only the $K\bar{K}$ on shell singularity. The final loop function that appeared in the calculation of Ref. \cite{wuzou}, containing the three propagators, was divergent and needed to be regularized. In Ref. \cite{wuzou} Wu \textit{et al.} used an unknown cutoff or form factor to implement convergence, but then the decay rates were dependent on an unknown parameter. The problem was solved in Ref. \cite{acetieta} by means of the chiral unitary approach and it was shown that the triangular loop could be regularized with the same cutoff used in the meson-meson interaction problem, which is a parameter fitted to the scattering data that naturally appears in the calculation. This made possible to evaluate the ratio of the decay rates for $\eta^{\prime}\rightarrow\pi^0\pi^0\eta$ and $\eta^{\prime}\rightarrow\pi^0\pi^+\pi^-$ and the shapes of the invariant mass distributions, which were in good agreement with experiment. 

In this work we will apply the same mechanism of Refs. \cite{wuzou, acetieta} to the reactions $f_1(1285)\rightarrow\pi^0\pi^0\eta$ and $f_1(1285)\rightarrow\pi^0\pi^+\pi^-$. We follow the steps of Ref. \cite{acetieta} and evaluate the branching ratio for the decay of the $f_1(1285)$ to $a_0\pi$, excluding the  $a_0(980)$ decay to $K\bar{K}$, in order to compare our result with the value reported in the PDG, with the aim to give further support to the basic idea about the nature of the $f_1(1285)$, $a_0(980)$ and $f_0(980)$ resonances.

%*******************************************************************\\
%CONFRONTO TRA GRANDEZZE CALCOLATE CON QUESTO METODO E DATI SPERIMENTALI, SE LO STUDIO DEL PROCESSO CON QUESTO MECCANISMO HA ESITO POSITIVO, VUOL DIRE CHE ABBIAMO SUPPORTO PER L'IDEA CHE QUESTE RISONANZE SIANO DINAMICAMENTE GENERATE.\\
%**********************************************************************
%***************************************************************************
\section{Formalism}
\label{formalism}

We want to study the  decay of the $f_1(1285)$ to $a_0\pi$, excluding the  $a_0(980)$ decay to $K\bar{K}$, which according to the PDG \cite{PDG}, makes up for a sizeable fraction of the total $f_1(1285)$ decay width of $(36\pm 7)\%$. In addition, we will simultaneously study the reactions $f_1(1285)\rightarrow a_0(980)\pi^0$ and $f_1(1285)\rightarrow f_0(980)\pi^0$ to make an estimate of the amount of $a_0(980)-f_0(980)$ mixing. 

\begin{figure}[!ht]
\includegraphics[width=15cm,height=9cm]{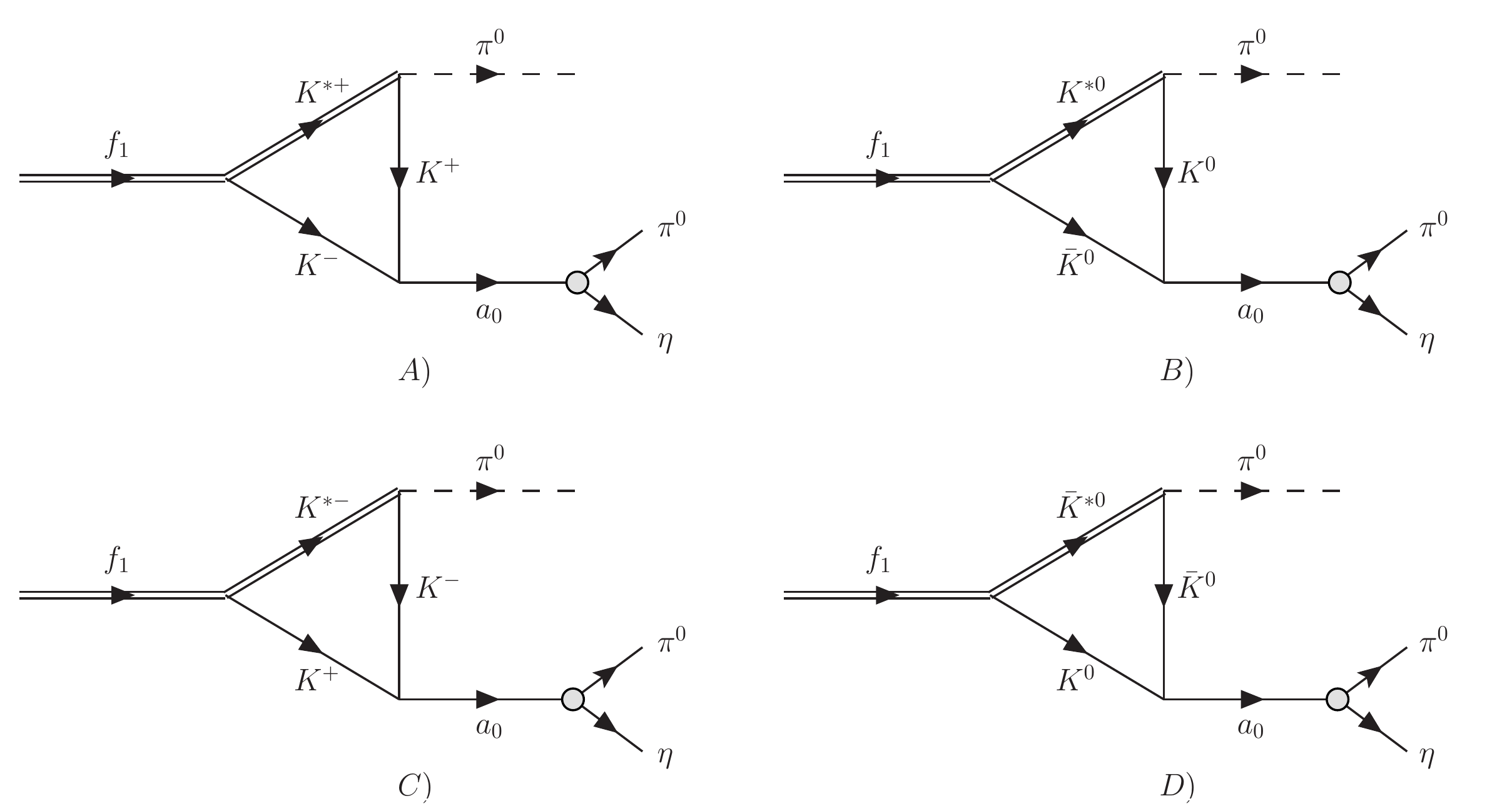}
\caption{Diagrams representing the process $f_1(1285)\rightarrow \pi^0\eta\  (\pi^+\pi^-)$.}
\label{fig:diagrams}
\end{figure}

Once the dynamically generated picture for the $f_1(1285)$, $a_0(980)$ and $f_0(980)$ is assumed, the process can be described by means of the triangular mechanism contained in the four diagrams of Fig. \ref{fig:diagrams}, consisting of the $f_1(1285)$ decay to $K^*\bar{K}$, the successive decay of the $K^*$ into $K\pi$ and the rescattering of the $K\bar{K}$ pair leading to $\pi^0\eta$ or $\pi^+\pi^-$ in the final state, which will proceed via the $a_0(980)$ or $f_0(980)$ resonances accounted for by the $K\bar{K}\rightarrow\pi^0\eta$ and $K\bar{K}\rightarrow\pi^+\pi^-$ amplitudes, respectively.

%***************************************************************************

\subsection{The structure of the vertices}
\label{vertices}
\begin{figure}[!ht]
\includegraphics[width=11cm,height=6.5cm]{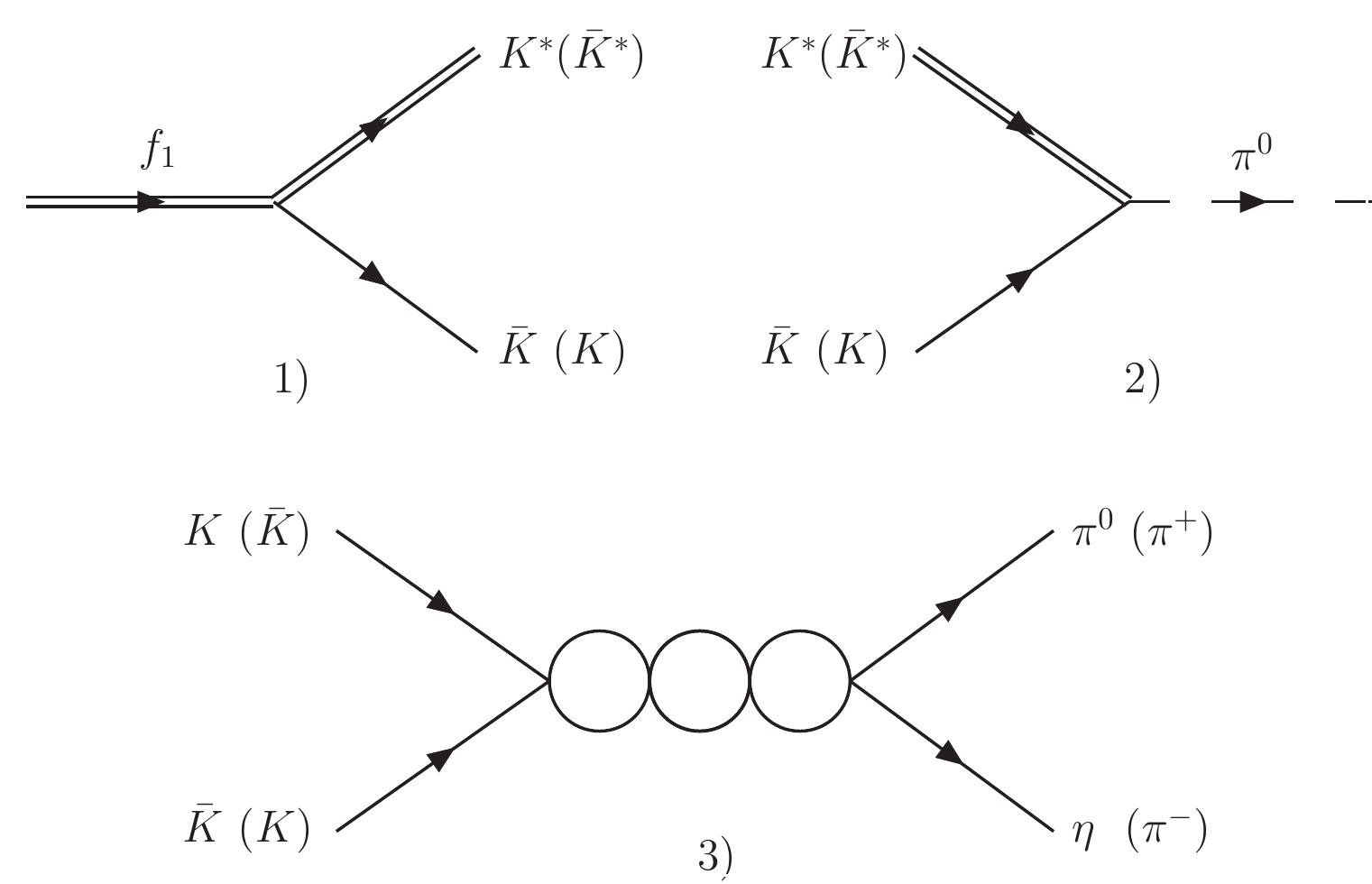}
\caption{Vertices involved in the decay of the $f_1(1285)$ to $\pi^0\eta$ or $\pi^+\pi^-$, the $f_1(1285)\rightarrow K^*\bar{K}$ 1) and the $VPP$ vertex 2), and amplitude for $a_0(980)$ and $f_0(980)$ production 3).}
\label{fig:vertices}
\end{figure}

In order to evaluate the amplitudes of the diagrams in Fig.\ref{fig:diagrams}, we need the structure of the two vertices involved plus the $K\bar{K}\rightarrow\pi^-\pi^+(\pi^0\eta)$ amplitude, shown in Fig. \ref{fig:vertices}. 

As mentioned before, in Ref. \cite{f1luis}, the $f_1(1285)$ results as dynamically generated from the interaction of $K^*\bar{K}-cc$. We can write the vertex 1) of Fig. \ref{fig:vertices} as
\begin{equation}
\label{eq:vertex1}
-it_{1}=-i\,g_{f_1}\,C_1\epsilon^{\mu}\epsilon_{\mu}^{\prime}\ ,
\end{equation}
where $\epsilon$ is the polarization vector of the $f_1$ and $\epsilon^{\prime}$ is the polarization vector of the $K^{*}\ (\bar{K}^*)$. 

The coupling $g_{f_1}$ of the $f_1$ to the $K^{*}\bar{K}$ channel is obtained from the residue in the pole of the scattering amplitude for $K^*\bar{K}-cc$ in $I=0$, which, in general, close to a pole and in the case of a single channel, can be written as
\begin{equation}
T\simeq\frac{g_R^2}{s-s_R}\ ,
\end{equation}
where $s_R$ is the position of the resonance in the complex energy plane and $g_R$ is the coupling to the channel. The scattering amplitude $T$ is obtained using the Bethe-Salpeter equation
\begin{equation}
T=[1-VG]^{-1}\ V\ ,
\label{eq:BS}
\end{equation}
with the potential $V$ taken from Ref. \cite{f1luis}. The $G$ function in Eq. \eqref{eq:BS} is the loop function for the propagators of the intermediate particles
\begin{equation}
\label{eq:G}
G(P^2)=\int\frac{d^4q}{(2\pi)^4}\frac{1}{q^2-m_1^2+i\epsilon}\ \frac{1}{(P-q)^2-m_2^2+i\epsilon}\ ,
\end{equation}
where $P$ the total four-momentum ($P^2=s$) and $m_1$, $m_2$ the masses of the particles in the considered channel. After the regularization by means of a cutoff \cite{npa} we get
\begin{equation}
\label{eq:G2}
G(s)=\int_{|\vec{q}\,|<q_{max}}\frac{d^3q}{(2\pi)^3}\frac{\omega_1+\omega_2}{2\omega_1\omega_2}\ \frac{1}{s-(\omega_1+\omega_2)^2+i\epsilon}\ ,
\end{equation}
with $\omega_i=\sqrt{|\vec{q}\,|+m_i^2}$.
We obtain a good description of the $f_1(1285)$ using a cutoff of about $1$\gev, as in Ref. \cite{f1luis}.

The factors $C_1$ in Eq. \eqref{eq:vertex1} are due to the fact that the $f_1(1285)$ couples to the $I=0$, $C=+$, $G=+$ combination of $K^{*}\bar{K}$ mesons, which is represented by the state
\begin{equation}
\label{eq:stateKKs}
\frac{1}{\sqrt{2}}(K^*\bar{K}-\bar{K}^*K)=-\frac{1}{2}(K^{*+}K^-+K^{*0}\bar{K}^0-K^{*-}K^+-\bar{K}^{*0}K^0)\ ,
\end{equation}
where the convention $CK^*=-K^*$ is taken, which is consistent with the standard chiral Lagrangians.
The different values of $C_1$ for each diagram of Fig. \ref{fig:diagrams}, corresponding to the weights of the charged and neutral components in the wave function of Eq. \eqref{eq:stateKKs}, are listed in the second column of Table \ref{tab:factors}.

The structure of the vertices of type 2) can be derived using the hidden gauge symmetry Lagrangian describing the $VPP$ interaction \cite{hidden1, hidden2, hidden3, hidden4}, given by
\begin{equation}
\label{eq:lagrangianhg}
\mathcal{L}_{PPV}=-ig\ \langle V^\mu [P,\partial_\mu P]\rangle\ ,
\end{equation}
where the symbol $\langle\rangle$ stands for the trace in $SU(3)$ and $g=\frac{m_V}{2f}$, with $m_V\simeq m_{ \rho}$ and $f=93$\mev\ the pion decay constant.

\begin{figure}[!ht]
\includegraphics[width=9cm,height=5cm]{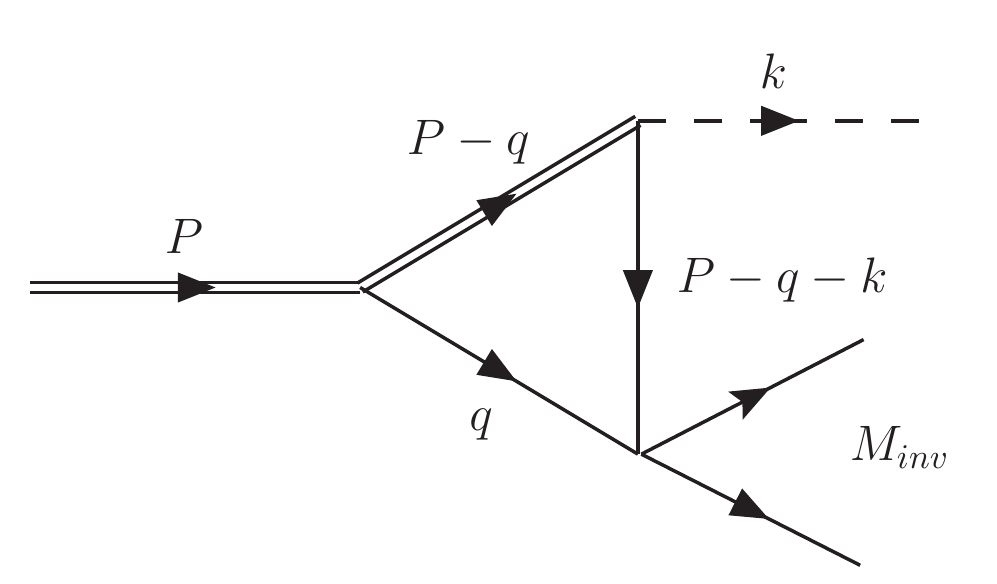}
\caption{Momenta assignment for the decay process.}
\label{fig:momenta}
\end{figure}

The P matrix in Eq. \eqref{eq:lagrangianhg} contains the nonet of the pseudoscalar mesons written in the physical basis in which $\eta$, $\eta'$ mixing is considered \cite{Bramon:1994pq},
\begin{equation}
P=\left(
\begin{array}{ccc}
\frac{\eta}{\sqrt{3}}+\frac{\eta'}{\sqrt{6}}+\frac{\pi^0}{\sqrt{2}} & \pi^+ & K^+\\
\pi^- &\frac{\eta}{\sqrt{3}}+\frac{\eta'}{\sqrt{6}}-\frac{\pi^0}{\sqrt{2}} & K^{0}\\
K^{-} & \bar{K}^{0} &-\frac{\eta}{\sqrt{3}}+\sqrt{\frac{2}{3}}\eta'\\
\end{array}
\right)\ ,
\label{eq:pfields}
\end{equation}
while the $V$ matrix contains the nonet of vector mesons,
\begin{equation}
V_\mu=\left(
\begin{array}{ccc}
\frac{\omega}{\sqrt{2}}+\frac{\rho^0}{\sqrt{2}} & \rho^+ & K^{*+}\\
\rho^- &\frac{\omega}{\sqrt{2}}-\frac{\rho^0}{\sqrt{2}} & K^{*0}\\
K^{*-} & \bar{K}^{*0} &\phi\\
\end{array}
\right)_\mu\ .
\label{eq:vfields}
\end{equation}
Thus, the amplitude of the vertex can be written as
\begin{equation}
-it_2=i\,g\,C_{2}\,(2k-P+q)_{\mu}\epsilon^{\prime\mu}\ ,
\label{eq:VPPampli}
\end{equation}
where the factors $C_2$ for each diagram in Fig. \ref{fig:diagrams} are shown in the third column of Table \ref{tab:factors}. The momenta in Eq. \eqref{eq:VPPampli} are assigned as shown in Fig. \ref{fig:momenta}.

\begin{table}[ tp ]%
\begin{tabular}{c||c|c|c}
\hline %
Diagram &  $\phantom{-}\phantom{-}C_{1}\phantom{-}\phantom{-}$ & $\phantom{-}\phantom{-}C_{2}\phantom{-}\phantom{-}$ & $\phantom{-}\phantom{-}C\phantom{-}\phantom{-}$\\\toprule %
A) & $-\frac{1}{2}$ & $\phantom{-}\frac{1}{\sqrt{2}}$ & $-\frac{1}{2\sqrt{2}}$ \\
B) & $-\frac{1}{2}$ & $-\frac{1}{\sqrt{2}}$ & $\phantom{-}\frac{1}{2\sqrt{2}}$ \\
C) & $\phantom{-}\frac{1}{2}$ & $-\frac{1}{\sqrt{2}}$ & $-\frac{1}{2\sqrt{2}}$ \\ 
D) & $\phantom{-}\frac{1}{2}$ & $\phantom{-}\frac{1}{\sqrt{2}}$ & $\phantom{-}\frac{1}{2\sqrt{2}}$ \\\hline
\end{tabular}
\caption{Factor $C_1$ and $C_2$ for the vertices 1) and 2) in Fig. \ref{fig:vertices} and for the four diagrams in Fig. \ref{fig:diagrams} in the second and third column, respectively. The values of their product, $C$, for the four different diagrams are listed in the fourth column.}
\label{tab:factors}\centering %
\end{table}

The $K\bar{K}\rightarrow \pi^0 \eta (\pi^+\pi^-)$ amplitude in Fig. \ref{fig:vertices} corresponds to the mechanism for the production of either $\pi^0\eta$ or $\pi^+$ $\pi^-$ in the final state. The rescattering of the $K\bar{K}$ pair dynamically generates in coupled channels the $a_0(980)$ and $f_0(980)$ resonances \cite{npa}, leading, respectively, to the $\pi^0\eta$ and $\pi^+\pi^-$ pair production. However, we must keep in mind that the $f_1(1285)$ is an $I=0$ object. If isospin symmetry were an exact symmetry (or if the kaons had the same mass), the process would only go via $a_0$ production, which is $I=1$, and this would prevent finding the $\pi^+\pi^-$ pair in the final state in $s$-wave (the $\rho^0$ in $p$-wave is forbidden by $C$-parity conservation). When the physical masses of the kaons are considered, we have an isospin breaking effect that leads to the production of the $f_0$ and then of the $\pi^+\pi^-$ pair.

We will write, for simplicity, this vertex as
\begin{equation}
\label{eq:vertex3}
-it_3=-it_{if}\ ,
\end{equation}
where $t_{if}$ is the $if$ element of the $5\times 5$ scattering matrix $t$ for the channels $K^+K^-$ (1), $K^0\bar{K}^0$ (2), $\pi^0\eta$ (3), $\pi^+\pi^-$ (4) and $\pi^0\pi^0$ (5) \cite{npa}. We have $i=1$ for the diagrams A) and C) and $i=2$ for the diagrams B) and D) of Fig. \ref{fig:diagrams}, while the index $f$ stands for the channel $3$ or $4$ depending on the meson pair in the final state. The $t$ matrix is obtained again using the Bethe-Salpeter equation, Eq. \eqref{eq:BS}\footnote{Note that in order to get the $t$ matrix, which sums over intermediate states, the unitary normalization of the $\pi^0\pi^0$ state implementing an extra $\frac{1}{\sqrt{2}}$ factor in the normalization of the state with identical particles, is done in the calculation \cite{npa}.}. 

The $G$ function in Eq. \eqref{eq:BS} is the diagonal loop function matrix for the intermediate states. As in the case of the $f_1$ production, the loops are regularized using the cutoff method. 

A good description of the $a_0(980)$ and $f_0(980)$ is obtained using for the loop functions a cutoff around $900$\mev. We will see in the next section that this parameter enters the evaluation of the loop integral in the diagrams of Fig. \ref{fig:diagrams}. 

%***************************************************************************

\subsection{The triangular loop}
\label{loop}

Putting together Eqs. \eqref{eq:vertex1}, \eqref{eq:VPPampli} and \eqref{eq:vertex3}, we can explicitly write the total amplitude for each one of the diagrams in Fig. \ref{fig:diagrams} as 
\begin{equation}
\label{eq:ampl1}
\begin{split}
-it&=-i\,C_1\,g_{f_1}\,i\,C_2\,g\,\int\frac{d^4q}{(2\pi)^4}(2k-P+q)_{\mu}\epsilon^{\prime\mu}\,\epsilon_{\alpha}\epsilon^{\prime\alpha}\frac{i}{q^2-m_K^2+i\epsilon}\\
&\times\frac{i}{(P-k-q)^2-m_K^2+i\epsilon}\,\frac{1}{2\omega^{*}(q)}\,\frac{i}{P^0-q^0-\omega^{*}(q)+i\epsilon}\,(-it_{if})\ ,
\end{split}
\end{equation}
where $\omega^{*}(q)=\sqrt{\vec{q}^{\ 2}+m_{K^*}^2}$ is the $K^*$ energy. 
In Eq.\eqref{eq:ampl1} only the positive energy part of the $K^*$ propagator $i[(P^0-q^0-\omega^{*})2\omega^{*}]^{-1}$ is taken, which is a good approximation given the large mass of the $K^*$.

We assume we are dealing with small three-momenta compared to the masses of the particles involved. This means that only the spatial components of the polarization vector of the $K^*$ are non vanishing,
\begin{equation}
\epsilon^{\prime0}=\frac{|\vec{P}-\vec{q}\,|}{m_{K^*}}=\frac{|\vec{q}\,|}{m_{K^*}}\sim0\ ,
\end{equation}
and that the completeness relation for the polarization vectors, which reads
\begin{equation}
\label{eq:epseps}
\sum_ {pol}\epsilon^{\prime}_{\mu}\epsilon^{\prime}_{\alpha}=-g_{\mu\alpha}+\frac{(P-q)_{\mu}(P-q)_{\alpha}}{m^2_{K^{*}}}\ ,
\end{equation}
can now be written as
\begin{equation}
\label{eq:epseps2}
\sum_ {pol}\epsilon^{\prime}_{\mu}\epsilon^{\prime}_{\alpha}\simeq \sum_ {pol}\epsilon^{\prime}_{i}\epsilon^{\prime}_{j}=\delta_{ij}\ ;\ \ \ \ \ \ \mu=i,\ \ \alpha=j;\ \ i,j=1,2,3\ .
\end{equation}
Using Eq. \eqref{eq:epseps2}, the amplitude reduces to
\begin{equation}
\label{eq:ampl2}
\begin{split}
t&=-i\,C_1\,C_2\,g_{f_1}\,g\,\int\frac{d^4q}{(2\pi)^4}(2\vec{k}+\vec{q})\cdot\vec{\epsilon}\,\frac{1}{q^2-m_K^2+i\epsilon}\,\frac{1}{(P-k-q)^2-m_K^2+i\epsilon}\\&\times\frac{1}{2\omega^{*}(q)}\,\frac{1}{P^0-q^0-\omega^{*}(q)+i\epsilon}\,t_{if}\ .
\end{split}\ .
\end{equation}

We can further simplify Eq. \eqref{eq:ampl2} writing it as
\begin{equation}
\label{eq:ampl3}
t=\,C\,g_{f_1}\,g\,\vec{\epsilon}\cdot\vec{k}\,(2I_1+I_2)\,t_{if}=\tilde{t}\,\vec{\epsilon}\cdot\vec{k}\,t_{if}\ ,
\end{equation}
where $I_1$ and $I_2$ are defined as
\begin{equation}
\begin{split}
\label{eq:int1d4}
&I_1=-i\ \int\frac{d^4q}{(2\pi)^4}\frac{1}{q^2-m_K^2+i\epsilon}\,\frac{1}{(P-k-q)^2-m_K^2+i\epsilon}\ \frac{1}{2\omega^{*}(q)}\,\frac{1}{P^0-q^0-\omega^{*}(q)+i\epsilon}\ ,\\
&I_2=-i\ \int\frac{d^4q}{(2\pi)^4}\frac{1}{q^2-m_K^2+i\epsilon}\,\frac{\vec{k}\cdot\vec{q}/|\vec{k}\,|^2}{(P-k-q)^2-m_K^2+i\epsilon}\ \frac{1}{2\omega^{*}(q)}\,\frac{1}{P^0-q^0-\omega^{*}(q)+i\epsilon}\ .\\
\end{split}
\end{equation}
The constant $C$ is the product of $C_1$ and $C_2$ and its value depends on the diagram that we are considering, as shown in the last column of Table \ref{tab:factors}.

After analytically integrating Eqs. \eqref{eq:int1d4} in $dq^0$ using Cauchy's theorem, we obtain
\begin{equation}
\begin{split}
\label{eq:int1d3}
I_1&=-\int\frac{d^3q}{(2\pi)^3}\,\frac{1}{8\,\omega(q)\omega^{\prime}(q)\omega^*(q)}\,\frac{1}{k^0-\omega^{\prime}(q)-\omega^{*}(q)+i\epsilon}\,\frac{1}{P^0-\omega^*(q)-\omega(q)+i\epsilon}\\
&\times \frac{2P^0\omega(q)+2k^0\omega^{\prime}(q)-2(\omega(q)+\omega^{\prime}(q))(\omega(q)+\omega^{\prime}(q)+\omega^{*}(q))}{(P^0-\omega(q)-\omega^{\prime}(q)-k^0+i\epsilon)(P^0+\omega(q)+\omega^{\prime}(q)-k^0-i\epsilon)}\ ,
\end{split}
\end{equation}
\begin{equation}
\begin{split}
\label{eq:int2d3}
I_2&=-\int\frac{d^3q}{(2\pi)^3}\,\frac{\vec{k}\cdot\vec{q}/|\vec{k}\,|^2}{8\,\omega(q)\omega^{\prime}(q)\omega^*(q)}\,\frac{1}{k^0-\omega^{\prime}(q)-\omega^{*}(q)+i\epsilon}\,\frac{1}{P^0-\omega^*(q)-\omega(q)+i\epsilon}\\
&\times \frac{2P^0\omega(q)+2k^0\omega^{\prime}(q)-2(\omega(q)+\omega^{\prime}(q))(\omega(q)+\omega^{\prime}(q)+\omega^{*}(q))}{(P^0-\omega(q)-\omega^{\prime}(q)-k^0+i\epsilon)(P^0+\omega(q)+\omega^{\prime}(q)-k^0-i\epsilon)}\ ,
\end{split}
\end{equation}
where $\omega(q)=\sqrt{\vec{q}^{\ 2}+m_K^2}$ and $\omega^{\prime}(q)=\sqrt{(\vec{q}+\vec{k})^{2}+m_K^2}$ are the energies of the $K$ ($\bar{K}$) and $\bar{K}$ ($K$) in the loop respectively.

When performing numerically the integrations in $d^3q$ of Eqs. \eqref{eq:int1d3} and \eqref{eq:int2d3} we have to consider that the upper limit is naturally provided, as in Ref. \cite{acetieta}, by the chiral unitary approach. As we already mentioned in Section \ref{vertices}, the loop function $G$ used in meson meson scattering to generate the $a_0(980)$ and $f_0(980)$ is divergent and regularized  by a cutoff fitted to the experimental data. Using this same cutoff in the triangular loop is not only natural but also necessary if we remember that the implementation of a cutoff $\theta(q_{max}-|\vec{q}|)$ in the integration of $G$ is done in a quantum mechanical formulation which, in the case of $s$-waves, makes use of a potential of the form
\begin{equation}
\label{eq:cutoffpotential}
V(\vec{q},\vec{q}^{\ \prime})=v\,\theta(q_{max}-|\vec{q}|)\,\theta(q_{max}-|\vec{q}^{\ \prime}|),
\end{equation}
which leads to 
\begin{equation}
\label{eq:cutoffamplitude}
t(\vec{q},\vec{q}^{\ \prime})=t\,\theta(q_{max}-|\vec{q}|)\,\theta(q_{max}-|\vec{q}^{\ \prime}|).
\end{equation}
This means that the cutoff $q_{max}$ appears automatically as the upper limit of the integrals of Eqs. \eqref{eq:int1d3} and \eqref{eq:int2d3} thanks to the $K\bar{K}\rightarrow PP$ potential used to dynamically generate the $a_0$ and $f_0$. Note, however, that the integrals in Eqs. \eqref{eq:int1d3} and \eqref{eq:int2d3} are already convergent without implementing $q_{max}$. The loop also should implement the cutoff used in the evaluation of the $f_1(1285)$ which was $q_{max}=1000$ MeV. Hence the use of $q_{max}=900$ MeV accounts for both cutoffs.

Proceeding with the evaluation of the total amplitude of the reactions $f_1(1285)\rightarrow \pi^0\pi^0\eta$ and $f_1(1285)\rightarrow \pi^0\pi^+\pi^-$, we have to take into account that the neutral and charged kaons have different physical masses. Thus, we define $\tilde{t}^{(+)}$ and $\tilde{t}^{(0)}$, corresponding to the quantity $\tilde{t}$ of Eq. \eqref{eq:ampl3} evaluated for the masses of $K^+$, $K^-$, $K^{*+}$, $K^{*-}$ (summing A and C of Fig. \ref{fig:diagrams}) and $K^0$, $\bar{K}^0$, $K^{*0}$, $\bar{K}^{*0}$ (summing B and D of Fig. \ref{fig:diagrams}) respectively. This allows us to write the total amplitudes of the processes as
\begin{equation}
\begin{split}
\label{eq:Ttotal}
&T_{\pi^0\eta}=(2\tilde{t}^{(+)}\,t_{K^+K^-\rightarrow\pi^0\eta}+2\tilde{t}^{(0)}\,t_{K^0\bar{K}^0\rightarrow\pi^0\eta})\,\vec{\epsilon}\cdot\vec{k}\ ,\\
&T_{\pi^+\pi^-}=(2\tilde{t}^{(+)}\,t_{K^+K^-\rightarrow\pi^+\pi^-}+2\tilde{t}^{(0)}\,t_{K^0\bar{K}^0\rightarrow\pi^+\pi^-})\,\vec{\epsilon}\cdot\vec{k}\ .
\end{split}
\end{equation}

From these last two equations, the role of the mass difference between neutral and charged kaons in the isospin symmetry breaking can be clearly understood. When equal masses for the kaons are taken, due to the fact that the global factor $C$ in $\tilde{t}$ has opposite sign in the charged and in the neutral case (see Table \ref{tab:factors}), we have that $\tilde{t}^{(0)}=-\tilde{t}^{(+)}$. Moreover, from Ref. \cite{npa} we know that 
\begin{equation}
\begin{split}
t_{K^+K^-\rightarrow\pi^0\eta}&=-t_{K^0\bar{K}^0\rightarrow\pi^0\eta}\ ,\\
t_{K^+K^-\rightarrow\pi^+\pi^-}&=t_{K^0\bar{K}^0\rightarrow\pi^+\pi^-}\ .
\end{split}
\end{equation}
This means that if the masses of the neutral and charged kaons were equal, the amplitude $T_{\pi^+\pi^-}$ would vanish, preventing the production of the $f_0(980)$ as intermediate state, which indeed is isospin forbidden. Since the isospin symmetry is not an exact symmetry, due to the mass difference, the decay can go via both $a_0$ and $f_0$ production, leading to the $\pi^0\eta$ and $\pi^+\pi^-$ pairs in the final state.

Given the structure of Eq. \eqref{eq:Ttotal} we define $T_{\pi^0\eta}$ and $T_{\pi^+\pi^-}$ as
\begin{equation}
\label{eq:Ttotal2}
\begin{split}
T_{\pi^0\eta}&=\tilde{T}_{\pi^0\eta}\,\vec{\epsilon}\cdot\vec{k}\ , \\
T_{\pi^+\pi^-}&=\tilde{T}_{\pi^+\pi^-}\,\vec{\epsilon}\cdot\vec{k}\ .
\end{split}
\end{equation}

\section{Results}
The invariant mass distribution is given by the formula
\cite{Xie:2014tma}
\begin{equation}
\label{eq:invmass}
\frac{d\Gamma}{dM_{inv}}=\frac{1}{(2\pi)^3}\,\frac{p_{\pi}\,|\vec{k}\ |}{4m_{f_1}^2}\,\frac{1}{2}\int_{-1}^{1}d\cos{\theta}\ \overline{\Sigma}|T|^2\ ,
\end{equation}
where the symbol $\overline{\Sigma}$ stands for the average over the polarizations of the $f_1(1285)$, $\theta$ is the angle between $\vec{k}$ and $\vec{\epsilon}$ and $M_{inv}$ is the invariant mass of the final interacting pair (see Fig. \ref{fig:momenta}). The momenta in Eq. \eqref{eq:invmass} are defined as
\begin{equation}
p_{\pi}=\frac{\lambda^{1/2}(M_{inv}^2,m_{\pi^0}^2,m_{\eta}^2)}{2M_{inv}}\ 
\label{eq:p1}
\end{equation}
in the case of $\pi^0\eta$ in the final state ($\pi^0$ momentum un the $\pi^0\eta$ rest frame) and
\begin{equation}
p_{\pi}=\frac{\lambda^{1/2}(M_{inv}^2,m_{\pi^+}^2,m_{\pi^-}^2)}{2M_{inv}}
\label{eq:p2}
\end{equation}
in the case of $\pi^+\pi^-$ ($\pi^+$ momentum un the $\pi^+\pi^-$ rest frame), while
\begin{equation}
|\vec{k}\,|=\frac{\lambda^{1/2}(M_{f_1}^2,m_{\pi^0}^2,M_{inv}^2)}{2m_{f_1}}\ 
\label{eq:k}
\end{equation}
is the momentum of the spectator $\pi^0$ in the reference frame in which the $f_{1}(1285)$ is at rest. The function $\lambda$ in Eqs. \eqref{eq:p1}, \eqref{eq:p2} and \eqref{eq:k} is the K\"{a}ll\'{e}n function.

Eq. \eqref{eq:invmass} can be rewritten, after performing the integration in $d\cos{\theta}$, as
\begin{equation}
\label{eq:invmass2}
\frac{d\Gamma}{dM_{inv}}=\frac{1}{(2\pi)^3}\,\frac{p_{\pi}\,|\vec{k}\ |^{3}}{4m_{f_1}^2}\,\frac{1}{3}|\,\tilde{T}|^2\ .
\end{equation}

\begin{figure}[ht!]
\includegraphics[width=12cm,height=6.5cm]{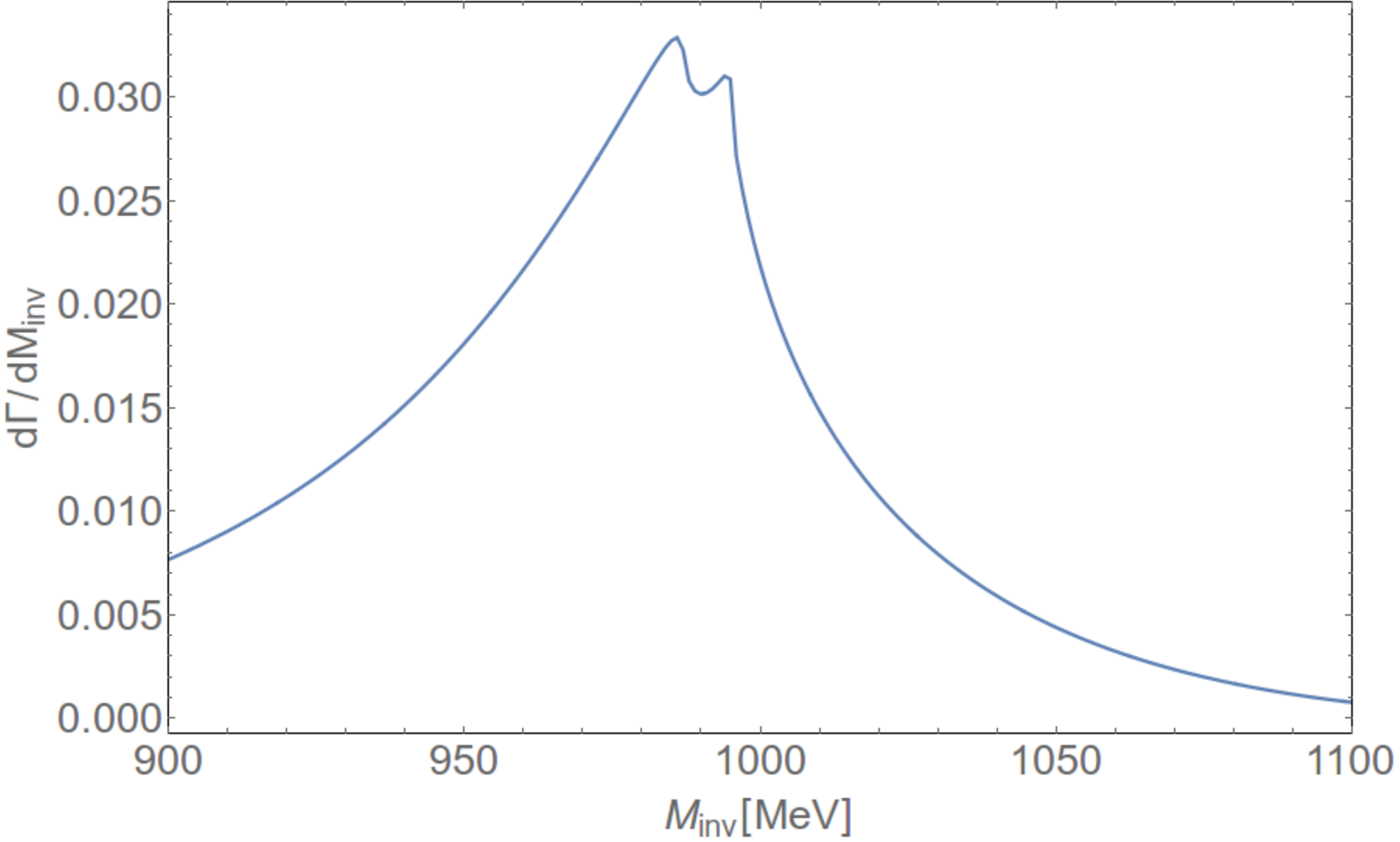}
\caption{$d\Gamma/dM_{inv}$ for $f_1(1285)\rightarrow\pi^0\pi^0\eta$ decay as a function of $M_{inv}$ in the $f_0(980)$ region.}
\label{fig:ratea0}
\end{figure}
\begin{figure}[ht!]
\includegraphics[width=12cm,height=6.5cm]{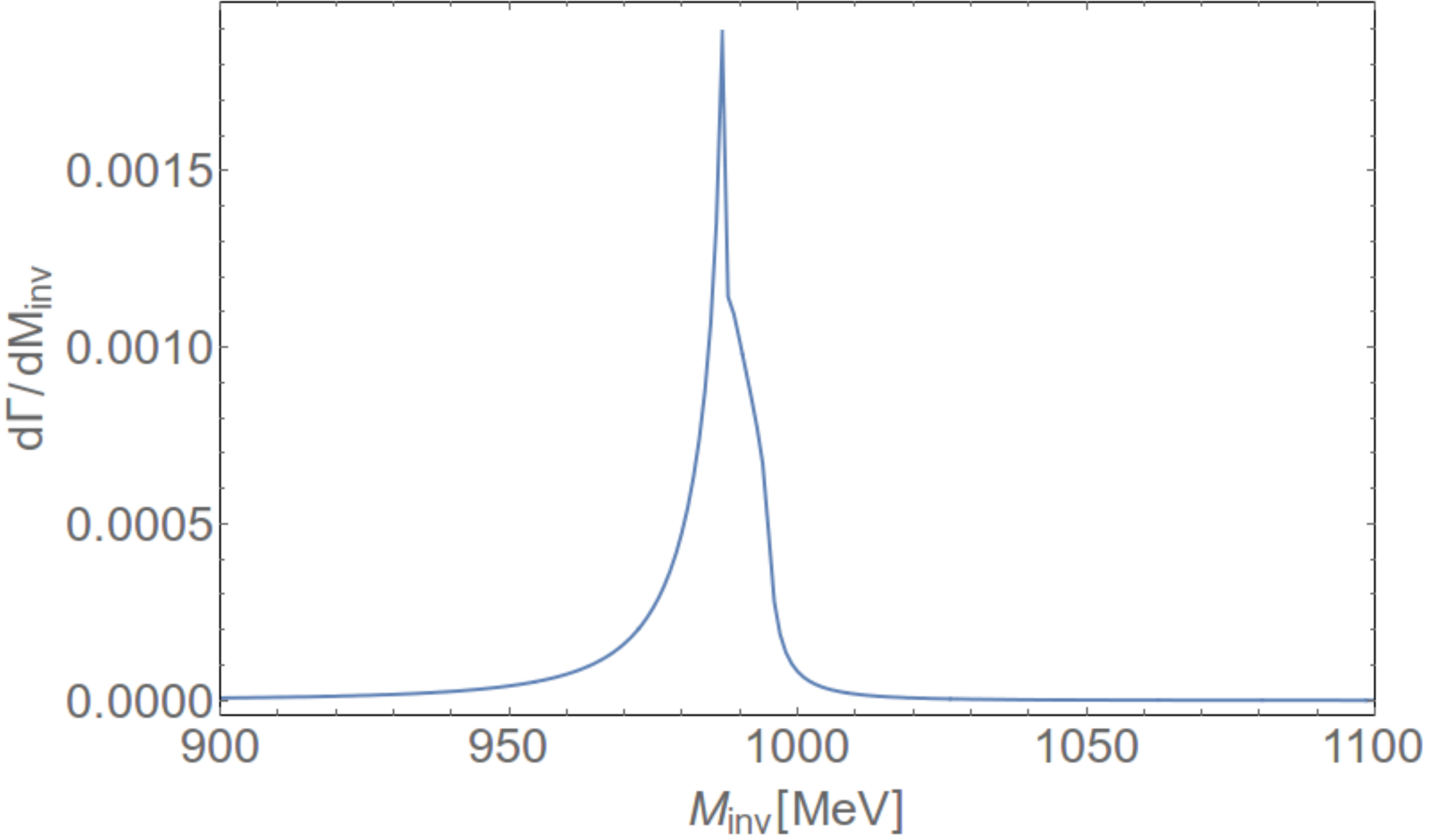}
\caption{$d\Gamma/dM_{inv}$ for $f_1(1285)\rightarrow\pi^0\pi^+\pi^-$ decay as a function of $M_{inv}$ in the $f_0(980)$ region.}
\label{fig:ratef0}
\end{figure}

The result for $d\Gamma/dM_{inv}$ for the  $f_1(1285)\rightarrow\pi^0\pi^0\eta$ case is shown in Fig. \ref{fig:ratea0} while in Fig. \ref{fig:ratef0} we have the same plot for  $f_1(1285)\rightarrow\pi^0\pi^+\pi^-$. For the same reason as in \cite{acetieta} we do not symmetrize the two $\pi^0$ in the $\pi^0\pi^0\eta$ final state, since the $a_0$ resonance is relatively narrow and the two $\pi^0$ have very distinct kinematics. 

\begin{figure}[ht!]
\includegraphics[width=7.9cm,height=4.5cm]{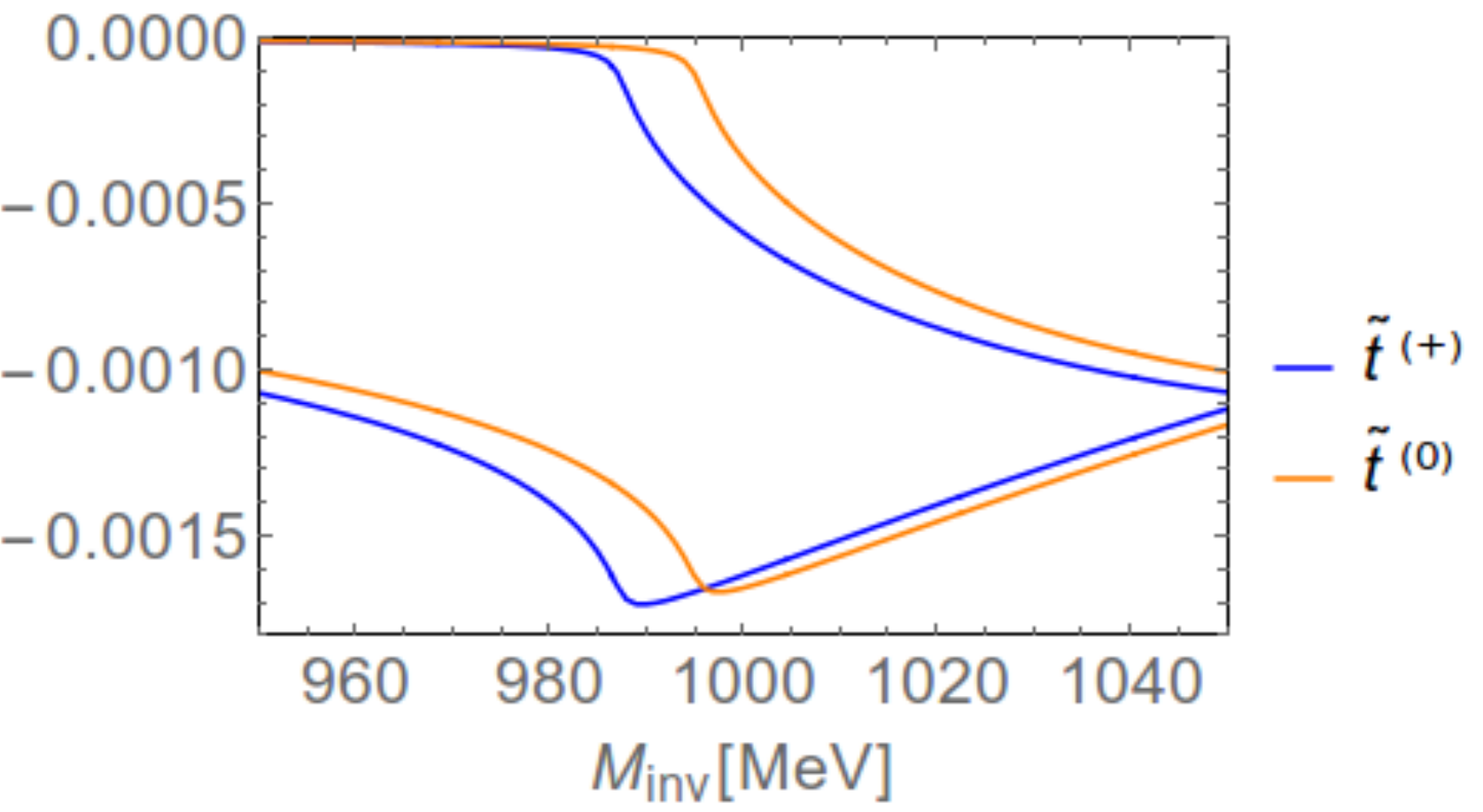}
\caption{Lines upper side: higher Im$[\tilde{t}^{(0)}]$, lower Im$[\tilde{t}^{(+)}]$; lines lower side: higher Re$[\tilde{t}^{(0)}]$, lower Re$[\tilde{t}^{(+)}]$.}
\label{fig:loops}
\end{figure}

For $\pi^+\pi^-$ in the final state we obtain, as in the case of Ref. \cite{acetieta}, a narrow peak around $980$\mev. The width of the peak is unusually small, around $10$\mev, in agreement with what was found experimentally in other reactions like the one studied by BES collaboration \cite{BESIII:2012aa}, and it appears exactly in the $f_0(980)$ region between the two thresholds of $K^*+K^-$ and $K^0\bar{K}^0$. This width is not the natural one of the $f_0$, which is around $60$\mev, and the shape we see in Fig. \ref{fig:ratef0} is different from the usual one seen in isospin allowed reactions. The reason for the peculiar features of this distribution can be found, as mentioned before, in the difference in the physical masses of neutral and charged kaons. This difference, as it can be seen in Fig. \ref{fig:loops}, is significant only in the region of energies around the two $K\bar{K}$ thresholds. Far from the thresholds the difference between the two loop integrals $\tilde{t}^{(+)}$ and $\tilde{t}^{(0)}$ becomes smaller, leading to the narrow shape of the invariant mass distribution for the $f_{0}(980)$, already observed in the case of the reaction $\eta(1405)\rightarrow\pi^0 f_{0}(980)$. 

\begin{figure}[ht!]
\includegraphics[width=12cm,height=6.5cm]{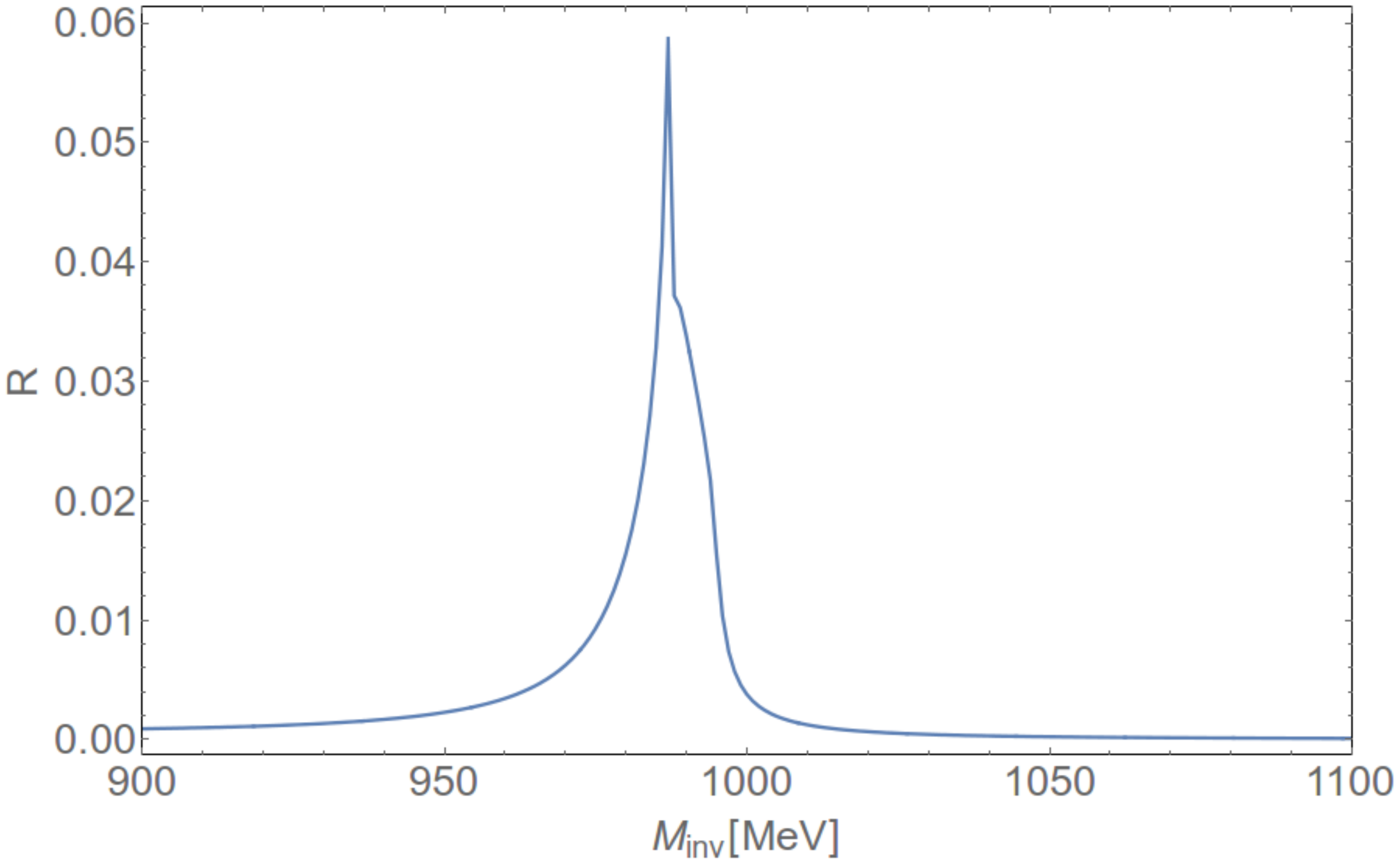}
\caption{Ratio $(\frac{d\Gamma}{dM_{inv}})_{\pi^+\pi^-}/(\frac{d\Gamma}{dM_{inv}})_{\pi^0\eta}$ as a function of $M_{inv}$.}
\label{fig:ratio}
\end{figure}

On the other hand, the signal is much wider in the case of $\pi^0\eta$ channel. We see in Fig. \ref{fig:ratea0} that the $a_0$ is produced with its normal width since the reaction is isospin allowed. Also the strength at the peak is much bigger than in the other case.
In Fig. \ref{fig:ratio} we show the ratio $(\frac{d\Gamma}{dM_{inv}})_{\pi^+\pi^-}/(\frac{d\Gamma}{dM_{inv}})_{\pi^0\eta}$ as a function of the invariant mass and we can see that the ratio of strength at the peak is of the order of $6\%$, ten times smaller than in Ref. \cite{acetieta} for the decay of the $\eta(1405)$. As discussed in \cite{acetieta}, we find once again that  there is not an absolute value of the $a_0-f_0$ mixing. It depends on the particular reaction but provides extra informations about the nature of the resonances and the reaction mechanism.

Now we proceed to the evaluation of the partial width for the decay mode of the $f_1(1285)$ to $a_0(980)\pi$. We want to compare our result with the experimental one reported in the PDG \cite{PDG}
\begin{equation}
BR(f_1\rightarrow a_0\pi)|_{exp}=(36\ \pm\ 7)\%\ ,
\label{eq:exp}
\end{equation}
ignoring the $a_0(989)\rightarrow K\bar{K}$ decay.

In order to do it, we need to take into account all the three possible final states, $a_0^+\pi^-$, $a_0^0\pi^0$, $a_0^-\pi^+$. However, the state of $I=0$ coming from the interaction of two $I=1$ particles is given by
\begin{equation}
|I=0,I_3=0\rangle=\frac{1}{\sqrt{3}}|1,1\rangle-\frac{1}{\sqrt{3}}|1,0\rangle+\frac{1}{\sqrt{3}}|1,-1\rangle\ ,
\end{equation} 
which means that the three final states appear with the same weight. Thus, we can restrict the calculation to the diagrams in Fig. \ref{fig:diagrams} and then multiply the result by a factor three to take into account the three charges. We find
\begin{equation}
\label{eq:width}
\Gamma_{a_0\pi}=3\ \int dM_{inv}\,\Big(\frac{d\Gamma}{dM_{inv}}\Big)_{\pi^0\eta}=7.6\ MeV\ ,
\end{equation}
with $\Big(\frac{d\Gamma}{dM_{inv}}\Big)_{\pi^0\eta}$ given by Eq. \eqref{eq:invmass2}, which corresponds to a branching ratio
\begin{equation}
BR(f_1\rightarrow a_0\pi)|_{th}\simeq\ 31\%\ ,
\label{eq:th}
\end{equation}
in good agreement with the experimental value.

We have made an estimate of the error on this result considering two possible sources of uncertainties. The first one is the cutoff in the meson-meson loop function used in Eq. \eqref{eq:BS} to generate the $f_1(1285)$, that we have taken around $1000$\mev. The value of the coupling $g_{f_1}$ used in the decay amplitude of Eq. \eqref{eq:ampl3} depends a bit on the cutoff. Changing its value by $\pm20$\mev, the resonance is still well reproduced and its position moved by only $10$\mev. This changes the value of the coupling $g_{f_1}$ by $2.5\%$, leading to the same uncertainty on the final result for $\Gamma_{a_0\pi}$. The other source of uncertainty is the cutoff $q_{max}$ used as upper limit in the loop integral of the decay. We make again a variation of $\pm 20$\mev, which moves the $a_0(980)$ and $f_0(980)$ peak in the scattering amplitude of $8$\mev. This induces a change in the value of $\Gamma_{a_0\pi}$ of $1.5\%$, which gives, summed to the uncertainty coming from the other source,

\begin{equation}
\begin{cases} 
\Gamma^{th}_{a_0\pi}=(7.6\pm 0.3)\ MeV\ ,\\
\\ 
BR(f_1\rightarrow a_0\pi)|_{th}=(31.4 \pm\ 1.2)\%\ . \end{cases}
\label{eq:thpluserr}
\end{equation}

The ratio of integrated strengths over the invariant mass in the region of the $a_0(980)$ and $f_0(980)$ for the reactions $f_1(1285)\rightarrow \pi^0\pi^0\eta$ and $f_1(1285)\rightarrow \pi^0\pi^+\pi^-$  gives
\begin{equation}
\label{eq:ratiointegrated}
\frac{\Gamma(\pi^0,\pi^+\pi^-)}{\Gamma(\pi^0,\pi^0\eta)}=0.82\times 10^{-2} ,
\end{equation}
\begin{equation}
\label{eq:ratiointegrated2}
\frac{\Gamma(\pi^0,f_0(980))}{\Gamma(\pi^0 a_0(980))}=1.28\times 10^{-2} ,
\end{equation}
where we have taken into account that the rate of $f_1(1285)\rightarrow\pi^0\pi^+\pi^-$ is twice the amount of $f_1(1285)\rightarrow\pi^0\pi^0\pi^0$.

This is much smaller than what was found for the $\eta(1405)$ decay, $\frac{\Gamma(\eta(1405)\rightarrow \pi^0f_0(980))}{\Gamma(\eta(1405)\rightarrow \pi^0a_0(980))}=18\times 10^{-2}$ \cite{BESIII:2012aa}, and twice as big as found for the $J/\psi$ decay, $\frac{\Gamma(J/\psi\rightarrow \phi f_0(980))}{\Gamma(J/\psi\rightarrow \phi a_0(980))}=0.6\times 10^{-2}$ \cite{Ablikim:2010aa}.

%**************************************************************************

\section{Conclusions}
We have evaluated the decay width of the $f_1(1285) \to \pi^0 \pi^0 \eta$, which shows a prominent peak in the $a_0(980)$ resonance region. We use the picture in which the  $f_1(1285)$ is dynamically generated from the vector-pseudoscalar interaction in the $K\bar{K^*} -c.c.$ channel and the $a_0(980)$ from the $\pi \eta, K \bar K$ channels. The mechanism for the decay consists in the triangular diagram of the  $f_1(1285)$ decaying into $K^* \bar K - c.c.$ and the $K^* (\bar K^*)$ decaying into $K (\bar K) \pi$, followed by the rescattering of the $K \bar K$ system. We find that the mechanism, which we can evaluate in absolute terms, provides a large branching fraction for the $f_1(1285)$ decay of about 30\%, in agreement with experiment. At the same time we evaluate the $f_1(1285) \to \pi^0 \pi^+ \pi^-$ decay rate, through the same mechanism, but with the $K \bar K$ scattering to produce $\pi^+ \pi^-$. This last process is isospin forbidden, and gives zero in our approach if we consider equal masses for the charged and neutral kaons. When physical masses are used, then isospin is slightly violated and we find a prominent peak, albeit with small intensity, in the $f_0(980)$ region. The width of this peak is found narrow, like in the $\eta(1405) \to \pi \pi \eta$ which was measured experimentally, and does not reflect the natural width of the $f_0(980)$ resonance but simply the region where the mass difference of the charged and neutral kaons is appreciable compared to the value of their masses. We find that the shape obtained is similar to the one found in the $\eta(1405) \to \pi^0 \pi^+ \pi^-$ and $J/\psi \to \phi \pi^0 \eta$, but the amount of isospin breaking is quite different to either reaction, showing once more that the concept of a universal mixing of the $f_0(980)$-$a_0(980)$ is not adequate, and the nature of these resonances as dynamically generated makes the isospin mixing very strongly dependent on the physical process. The measurement of the $f_1(1285) \to \pi^0 \pi^+ \pi^-$ decay and comparison with our predictions can serve to provide extra support for the picture in which the $f_1(1285)$, $f_0(980)$ and $a_0(980)$ resonances are dynamically generated. 

\section*{Acknowledgments}
This work is partly supported by the Spanish Ministerio de Economia y Competitividad
and European FEDER funds under the contract number FIS2011-28853-C02-01, and the
Generalitat Valenciana in the program Prometeo II, 2014/068. We acknowledge the support
of the European Community-Research Infrastructure Integrating Activity Study of Strongly
Interacting Matter (acronym HadronPhysics3, Grant Agreement n. 283286) under the Seventh
Framework Programme of EU. J. M. Dias would like to thank the Brazilian funding
agency FAPESP for the financial support.

\end{document}